\documentstyle[aps,preprint,prb,epsf]{revtex}
\begin{document}
\title{Distribution of Attraction Basins in a Family of Simple Glasses}
\author{P. Chandra$^1$ and L.B. Ioffe$^{2,3}$ \\
$^1$NEC Research Institute, 4 Independence Way, Princeton NJ 08540 \\
$^2$Landau Institute for Theoretical Physics, Moscow, RUSSIA \\
$^3$Department of Physics, Rutgers University, Piscataway, NJ 08855 \\
}
\maketitle

\begin{abstract}
We study the distribution of attraction basins as a function
of energy in simple glasses.
We find that it is always broad.
Furthermore we identify two types of glasses,
both with an exponentially large number of metastable
states.  In one type the largest attraction basin
is exponentially small, whereas in the other it
is polynomially small in the system size $N$.
If there exists a tuning parameter that connects
one regime with another, then these two
phases are separated by a critical point.
We discuss implications for optimization problems.
\end{abstract}

\pagebreak
\section{Introduction}

A complex system is one whose number of metastable
configurations, $N_s$,
scales exponentially with the number of its elements, $N$.
Naively one expects that an exponential number of searches is required
to find the optimal state.  In general the
identification of the ground-state in a complex system can be
mapped onto a hard combinatorial optimization problem.
\cite{Mezard87}
However there exist examples in nature, 
e.g. proteins,\cite{Wolynes97,Finkelstein97,Dill97,Dobson98} of complex
systems that find their ground-states on time-scales significantly
faster than $\tau \sim N_s$. A possible explanation for this
phenomenon
is that the associated total phase volume is not equally divided among
the metastable states.  More specifically, if a significant fraction
of the total phase volume belongs to the attraction basin of the 
optimal state, then a fast process leading to this ``greedy''
configuration
becomes feasible to implement.  A less stringent possibility is that
the optimal state can be located relatively quickly if it is connected
by a continuous path in the space of parameters to a state with a
large basin of attraction.  This is the underlying approach in
simulated annealing, an optimization algorithm that is very effective
for problems where the ground-state evolves continuously from the
paramagnetic configuration as a function of decreasing 
temperature.\cite{Kirkpatrick83}
However simulated annealing cannot be applied to systems where all the
metastable states appear at the same temperature.\cite{Bouchaud94,qannealing}
In this case an
open question is whether one can identify a parameter that connects
the state continuously to one with a large basin of attraction.  We
have addressed this issue in a family of simple glasses characterized
by a parameter $x$.  In particular we find that this parameter can
be increased continuously such that there exists a small subset of 
metastable states which attract the system with significant
probability $\cal{P}$ such that 
$\frac{1}{N} \ln \frac{1}{\cal{P}} \ll1$ 
in contrast to the generic situation 
$\frac{1}{N} \ln \frac{1}{\cal{P}}\sim 1$ in complex systems.

Here we study the basins of attraction in a family of $p$-spin
spherical models\cite{Crisanti92} characterized by the Hamiltonian
\begin{equation}
H = \sqrt{x} \sum_{i_1i_2} J_{i_1i_2} s_{i_1} s_{i_2} 
    + \sqrt{1-x} \sum_{i_1i_2i_3} J_{i_1i_2i_3} s_{i_1} s_{i_2}
    s_{i_3}
\label{H(x)}
\end{equation} 
where the constraint $\sum^N_{i=1} s_i^2 = 1$ is satisified
by the $N$ spins that are represented by real variables $s_i$.
$J_{i_1i_2}$
and $J_{i_1i_2i_3}$ are two- and three-spin infinite-range couplings
respectively; furthermore they are completely random of arbitrary
sign.  In this family of glasses, we find that
the configurational entropy, 
${\cal S}_c \equiv \ln N_s$, remains extensive $({\cal S}_c \propto N)$
for $0 \le x < 1$ whereas ${\cal S}_c = 0$ for precisely $x=1$.  This
limit corresponds to the $p=2$ disordered spherical model which has
one stable solution\cite{Kosterlitz76} and therefore an associated
basin of attraction that is large.
The other extreme parameter limit of (\ref{H(x)}), $x = 0$, corresponds
to the three-spin spherical model with an extensive number
of metastable states\cite{Crisanti95,Cavagna97} whose attraction
volumes are each an exponentially small fraction of the full
phase space.\cite{Barrat98}  As an
aside, 
we note that here we
will use the normalizations
\begin{eqnarray}
\langle J_{i_1i_2}^2 \rangle &=& \frac{1}{8N}\\
\langle J_{i_1i_2i_3}^2 \rangle &=& \frac{1}{36N^2}
\label{norm}
\end{eqnarray}
where the angular brackets refer to an average over disorder; these
expressions, (\ref{norm}), are slightly different than those
commonly found in the literature but are convenient for this
family of mixed models.

The $p$-spin random spherical models $(p > 2)$ are believed to be the
simplest models that possess the essential properties of a generic 
complex system.\cite{Gross84}  Aside from having an extensive complexity 
(${\cal S}_c \propto N$), they also exhibit history-dependance and aging
characteristic of experimental glasses.\cite{Bouchaud98}  
The simplicity of these
models arises from the long-range nature of the interactions,
a feature that makes them accessible to direct analytic treatment;
second, all metastable states appear at the same 
temperature and
are orthogonal in the thermodynamic limit\cite{Crisanti92,Kurchan93}
which simplifies the
dynamical
equations for $N_s \rightarrow \infty$.  Naturally the evolution of
a particular system with specific couplings must be studied
numerically.
However the physical properties of the system averaged over all
possible realizations of couplings can be studied analytically
by a set of integral-differential 
equations.\cite{Crisanti93,Cugliandolo93}  The latter describe the 
properties of typical metastable states where stochastic processes
will take the averaged system.  Here we show that the solution of
these equations implies that for $x=0$ ($p=3$ spherical model) the 
distribution of attraction basins as a function of energy is broad,
though the
attraction volume associated with the ground-state is still an
exponentially
small part of the full phase space.\cite{Barrat98}  
This regime persists up to finite
$x_c \sim \frac{1}{2}$.  However we find that as we continuously
tune $x$ to $x > x_c$ the complexity remains extensive (${\cal S}_c
\propto N$) for $x_c < x < 1$ but the ground-state acquires a large
basin of attraction that far exceeds the average and is a significant
fraction of the total phase volume.

We extract the distribution of attraction basins from the dynamical
equations of the mixed model (\ref{H(x)}) as a function of $x$.  More
specifically,
we relate the size of an attraction basin associated with a physical
state to the critical overlap, $q^*$, between this state and a
partially randomized one that still evolves back to it.  In
particular,
the limit $q^* \rightarrow 0$ corresponds to a basin of attraction
that occupies a significant portion of the available phase volume.
The critical overlap $q^*$ is found from the solutions of the
dynamical equations of (\ref{H(x)}).

The outline of this paper is as follows.  In Section II we discuss the
general approach taken here, in particular the determination of the
critical overlap as a function of energy, $q^*(E)$.  In Section III we
apply this approach to the pure $p=3$ random spherical model and
study the attraction basins as a function of energy.  Next
we turn to the mixed model (\ref{H(x)}) and determine the
distribution
of attraction basins for different values of $x$.  We find a parameter
regime, $1 > x > x_c$, where the mixed model has a typical state
with large trapping probability ($\frac{1}{N} \ln \frac{1}{\cal P} \ll
1$); furthermore in this regime the mixed model has marginally stable
states.  In Section IV we summarize our results
in a discussion, noting  that this conclusion is based on a study of
dynamical
equations that appear in a broad class of 
glassy models;\cite{Bouchaud96} in particular
they also describe glassy systems without quenched 
disorder.\cite{Kirkpatrick87,Parisi95,Franz95,Chandra95}  These
equations have also been proposed on phenomenological grounds for the
description of freezing in structural glasses; they are the so-called
mode-coupling equations.\cite{Gotze84}  Thus we expect that our result is more
general
that the specific mixed model defined above.

\section{The Approach}

The key step in our approach is to extract the probability, ${\cal
P}$, with which the typical state attracts the system from the
dynamical equations of the $p$-spin model.  This probability, ${\cal
P}$,
is equal to the ratio of the attraction basin of the physical
state, $W_B$, and the full volume of phase space,
$W_{PH}$,
so that ${\cal P} = \frac{W_B}{W_{PH}}$.  We take the typical state
as a reference point and parametrize arbitrary points in phase space
by their respective angles to this configuration.  In this framework 
\begin{equation}
W_{PH} = {\cal N}^{N-2} \int_0^\pi d\theta \ \sin^{N-2} \theta 
\label{wph}
\end{equation}
where ${\cal N}^{N-2}$ is the volume of the $(n-2)$-th dimensional unit
sphere.  Similarly the volume of the attraction basin of the reference
state is
\begin{equation}
W_B = {\cal N}^{N-2} \int_0^\pi d\theta \ P(\theta) \ \sin^{N-2} \theta 
\label{wb}
\end{equation}
where $P(\theta)$ is the probability that the state at angle $\theta$
belongs to the basin of attraction associated with the reference
configuration.

We note that for $N \gg 1$ the main contribution to $W_{PH}$ in 
(\ref{wb}) arises from $\sin \theta \equiv 1$, whereas that of $W_B$
comes from the largest possible $\sin \theta$ such that $P(\theta)$
is finite.  Therefore the behavior of $P(\theta)$ for 
$\theta \sim \frac{\pi}{2}$ is crucial for this discussion.  It is
therefore convenient to rewrite the expressions for $W_{PH}$ and $W_B$
using the parametrization $q = \cos \theta$ where $q$ is the overlap
between the typical state and that at angle $\theta$.  In this
notation,
assuming that $N \gg 1$, we find that
\begin{equation}
W_{PH} = {\cal N}^{N-2} \int_0^\infty dq \ e^{-\frac{N}{2}q^2}
\label{wph2}
\end{equation}
and
\begin{equation}
W_B  = {\cal N}^{N-2} \int_0^\infty dq \ P(q) e^{-\frac{N}{2}q^2}
\label{wb2}
\end{equation}
where we note that the main contribution to $W_{PH}$, displayed in 
(\ref{wph2}), arises from small $q \sim \frac{1}{\sqrt{N}}$.
There are two possible scenarios:

\noindent{\sl (i) $P(q) \equiv 0$ for $q < q^*$ (where $q^* > \frac{1}{\sqrt{N}}$).}

\noindent We note that this threshhold coincides with the previous definition
of $q^*$, the critical overlap beyond which a partially randomized
state evolves away from the reference state.  In this case
\begin{equation}
{\cal P} = \frac{W_B}{W_{PH}} \propto e^{-\frac{N}{2}(q^*)^2}
\label{Pth}
\end{equation}
and
\begin{equation}
\frac{1}{N} \ln \frac{1}{{\cal P}} \sim 1.
\label{lnPth}
\end{equation}

\noindent{\sl (ii) $P(q) = f(q)$ so that there is no threshhold (i.e. $q^*
\rightarrow 0$).} 

\noindent Then the probability that the reference state
attracts
the system is {\sl not} exponentially small and
\begin{equation}
\frac{1}{N} \ln \frac{1}{{\cal P}} \ll 1.
\label{lnPb}
\end{equation}
Therefore the size of the attraction basin associated with the typical
state is determined by the value of the critical overlap 
$q^* \equiv \cos \theta^*$.  We expect $q^*$ to have a distribution
of finite width; in this case states with the smallest value of
$q^*$ will have exponentially larger attraction basins than the
others.  Furthermore if $q^* \rightarrow 0$ then these states will
have basins of attraction that are a significant fraction of the
full phase volume.

We now discuss how to extract $P(q)$ and $q^*$ from the dynamical
equations of the family of $p$-spin spherical models.  These
equations determine the time-evolution of averaged correlation
$(D_{tt'} = \langle s(t)s(t')\rangle)$
and response 
$\left(G_{tt'} = \left\langle \frac{\partial s(t)}{\partial h(t')}
\right\rangle\right)$ functions for arbitrary sample history.   In
order to find $P(q)$, we consider the evolution of a state where at
time $t_0$ a fraction $1-q$ of the total spins is randomized so that
at time $t_0 + \epsilon$ the system is in a random state corresponding
to overlap $q$ with the state at $t_0$.  In terms of $D_{tt'}$ and
$G_{tt'}$
this randomization translates into a boundary condition
\begin{eqnarray}
D_{t_0 + \epsilon,t'} &=& (1-q) D_{t_0t'}\label{Db}\\
G_{t_0 + \epsilon,t'} &=& (1-q) G_{t_0t'}
\label{Gb}
\end{eqnarray}
where $t_0 > t'$.  The solution of the dynamical equations yields
$Q(q) \equiv \lim_{t\rightarrow \infty} D_{tt_0-\epsilon}(q)$,
the average overlap between a typical state and that which has evolved
from it in the manner described above.  This quantity can be
interpreted
in a simple way if all metastable states are orthogonal; in this case
$Q(q)$ is equal to $P(q)$, the probability that the system evolves
back to its original configuration after a fraction $(1-q)$ spins has
been randomized. This is indeed the situation for the family of
spherical spin models that we study here where there is only
one-step replica symmetry breaking as in the pure $p=3$
spherical model;\cite{Crisanti92} 
details of the replica solution for the mixed
models are presented in Appendix I.

We note that $Q(q)$ depends implicitly on the properties of the
typical
state at time $t_0$.  The dynamical equations with random initial
conditions yield $Q(q)$ averaged over all typical states.  This
average
is dominated by the states for which the combined basin of attraction
is maximal; for instance if states are characterized by their energy,
this quantity $N(E)W_B(E)$ is usually largest for states with the
highest energy which are still stable, namely states that are
marginally stable.  In order to probe states with different energies,
one needs to introduce different initial conditions via source terms
in the dynamical equations.\cite{Houghton83,Franz95b,Barrat96,Andrei99}  The solution of these modified equations
yields $Q(q,E)$, the overlap averaged over typical states of fixed
energy $E$.

The dynamical equations for the family of spherical models with
Hamiltonian (\ref{H(x)}) have the form
\begin{equation}
(a_{t_1} + {\partial}_{t_1}) D_{t_1t_2} - 2G_{t_1t_2} 
- \frac{\beta^2}{2} \int \Pi_{t_1t} G_{tt_2} dt
- \frac{\beta^2}{2} \int \Sigma_{t_1t} D_{tt_2} dt = {\cal S}_D
\label{Deq}
\end{equation}
and 
\begin{equation}
(a_{t_1} + {\partial}_{t_1}) G_{t_1t_2}  
- \frac{\beta^2}{2} \int \Sigma_{t_1t} D_{tt_2} dt = \delta(t_1 - t_2)
\label{Geq}
\end{equation}
where $\beta$ is the inverse temperature, $\Sigma$ and $\Pi$ are
self-energy terms, $a_t$ is determined implicitly by the condition
$D_{tt} = G_{tt} = 1$ and ${\cal S}_D$ is a source term that fixes
the initial energy.  For this mixed model with $N \gg 1$, we have
\begin{eqnarray}
\Sigma_{t_1t_2} & = & 2 (1-x) (GD)_{t_1t_2} + x G_{t_1t_2}
\label{Sigma}\\
\Pi_{t_1t_2} & = & (1-x)D^2_{t_1t_2} + xD_{t_1t_2}
\label{Pi}
\end{eqnarray}
and
\begin{equation}
{\cal S}_D = \frac{6\beta E_0}{2+x} 
\left( D^2_{t_10}D_{t_20}(1-x) + xD_{t_10}D_{t_20}\right)
\label{source}
\end{equation}
where $E_0$ is the energy of the initial configuration at $t=0$.  In
the two
self-energies, (\ref{Sigma}) and (\ref{Pi}), we recover known
results for the $p=2$ and $p=3$ spherical models for $x=1$ and $x=0$
respectively.\cite{Crisanti93,Cugliandolo93}  
The source term, (\ref{source}), is derived by 
introducing a term $\delta(H(t=0) - E_0)$ into the functional
integral for the stochastic dynamics, representing it as an additional
integral over a Lagrangian multiplier where the latter is 
determined
by
the initial energy $E_0$.
In order to obtain $P(x,q,E)$, we solve this
system of equations varying these three parameters.

\section{Results}

In Figure 1 we display the spin-spin correlation function,
$D_{tt'}$, of the $x=0$ $(p=3)$ spherical model after a fast
quench (i.e. with random initial conditions) as a reference
starting point for our subsequent discussion.  Apart from a narrow
range $t' \approx t$, this correlation function obeys a 
scaling\cite{Cugliandolo93}
form $D_{tt'} \sim \left( \frac{t'}{t}\right)^\gamma$.

In Fig. 2 we show the correlation function for the solution when
the system was partially randomized at $t_0 = \frac{t}{2}$ as
described
in (\ref{Db}) and (\ref{Gb}).
As an aside, we note that here and in what follows we
present results for $\frac{t}{t_0}=2$; we have checked that
they are weakly dependent on this ratio.
As expected, increased randomization
leads to a decreasing overlap between the state at $t_0$ and at
$t=2t_0$.  We also note that $D_{t_1t_2}$ shows power-law
aging behavior, $D_{t_1t_2} \approx \left( \frac{t_1}{t_2}\right)$,
for $t_1,t_2 < t_0$; for $t_1,t_2 > t_0$ the relaxation starts
again and $D_{t_1t_2} = d\left(\frac{(t_2 - t_0)}{(t_1 -
t_0)}\right)$.
These curves were determined numerically for finite $t_0$; as
$t_0\rightarrow \infty$ we expect that the limiting value of these
overlaps tends either to $Q_0$ or to zero slowly.  For example, a factor
of two increase in the overall time changes $D_{tt'}$ a little bit
as displayed in Fig. 2 by the dashed and full curves.  Therefore
a systematic finite-time analysis is necessary to determine
the value of $q^*$.  The dashed and full curves presented in Fig. 2
indicate that $q^*$ lies between $q_1$ and $q_2$ because the
plot for $q_1$ decreases with increasing $t_0$ whereas the opposite
is true for $q_2$.  We note that the slow decrease in $Q(q_1)$
with overall time can be understood as a finite-time effect; more 
specifically $Q(q_1)$ follows the same time evoluation as $D_{t0}$
for the reference dyanmics without randomization (see Fig. 1) such
that
\begin{equation}
\frac{\ln D_{t_0 + \epsilon,t_0}}{\ln t} = \frac{\ln D_{t,0}}{\ln t} = 
- \frac {1}{4}.
\label{ftscaling}
\end{equation}
In order to obtain $q^*$ more precisely, we consider the derivative
$\frac{dQ(q)}{d\ln t}$. We determine $q^*$ from the equation
$\frac{dQ(q)}{d\ln t} = 0$ where we check that the value of $q^*$
obtained
in this fashion is not dependent on the overall measuring time.
We note that the result, namely that $q^*(E=0)$ is finite for $x=0$
($p=3$ spherical model), is consistent with the conclusions
of an earlier study of this model.\cite{Barrat98}

Until now, we have considered solutions to the dynamical
equations, (\ref{Deq}) and (\ref{Geq}), with random
initial conditions; as we have discussed earlier these
probe high energy states that are marginally stable, a feature
that is responsible for their power-law evolution.  We now
turn to lower energy states as shown in Fig. 3.  In order
to access them we must fix our initial energy to 
be $E_0 < E_C$.  In Fig. 2 we display typical spin-spin correlation
functions in this energy regime.  The red curve indicates
clearly that for sufficiently large $q$ the system recovers
its state at $t_0$.  For smaller $q$ the state at $t_0 + \epsilon$
evolves away from its reference state (see Fig. 4) in a manner
similar to that of $D_{tt'}$ with completely random initial conditions
(see fig. 1).  Despite this qualitatively different behavior
for large and small $q$, $q^*$ must be determined by the same
finite-time
scaling that was discussed earlier.  Such an analysis indicates
that $q^*(x=0,E)$ remains finite for all energies.  Therefore the
basins of attraction in the $p=3$ spherical model increase with
decreasing energy, but always remain exponentially small
compared with the full phase volume.\cite{Barrat98}

In weakly frustrated systems where the number of metastable states
is subexponential, some basins of attraction must be large.
An example of such a system is the $p=2$ 
spherical model.\cite{Kosterlitz76}  An
interesting
question is whether it is possible to have some large basins of
attraction
but an exponential total number of states.  We expect an exponential
number
of states in a mixed $p=2$ and $p=3$ spherical model, and therefore
study a family of such systems to see whether they ever
acquire typical states with large basins of attraction.

We have checked that there is one-step replica-symmetry
breaking for $0 \le x \le 1$, and details are presented
in Appendix I.  As a result, we know that all metastable
states appear at $T=T_c$ and that there is no further
subdivision of states at lower temperatures.  We
can therefore perform an enumeration of these configurations
at $T=0$.
We have verified by direct computation that the logarithm of the
number of states for this mixed model is
\begin{equation}
{\cal S}_C \equiv \ln {\cal N}(x) = 
\frac{N}{2} 
\left(2 + \ln (2-x) - \frac {24 (2 - x)}{(x+2)(3-x)^2}
\right)
\label{S(x)}
\end{equation}
which is zero only at $x=1$ ($p=2$) where $x$ is a mixing parameter
as defined in the Hamiltonian (\ref{H(x)}).  Details of the
calculation
that yields (\ref{S(x)}) are given in Appendix II. 

We repeated the numerical analysis outlined above for values
of $x$ such that $0 < x < 1$.  Our results for $q^*$ are summarized
in Fig. 5.  As shown there, for $x=0.3$ all basins of attraction
remain exponentially smaller than the full phase volume.  However
for $x=0.6$, 
the critical overlap $q^*(E)$ is zero at $E_0$ indicating that states of
energy $E_0$ have basins of attraction that occupy a significant
fraction of the full phase volume.

Now we discuss possible weak points in this argument.  We have assumed
the exact orthogonality of the metastable states which is only true to
order $\frac{1}{\sqrt{N}}$.  This might lead to 
$q^* \sim \frac{1}{\sqrt{N}}$
instead of $q^* = 0$ for the states with large attraction basins.
This
correction would result in ${\cal P} \sim N^{-\alpha}$
with $\alpha$ of order unity.  Another weak point in the argument
might be the effect of finite-size corrections to the
equations (11)-(17) which were originally derived in the thermodynamic
limit.  For the $p=3$ spherical model we have derived the subleading
terms in $\frac{1}{N}$ which modify the expressions for the
self-energies $\Sigma$ and $\Pi$, included them in equations
({\ref{Deq}) and ({\ref{Geq}), and have checked that their
effects are perturbative.  Furthermore, because the self-energy
scales as $D^3$, higher-order terms in $\frac{1}{N}$, e.g. terms
of order O($\frac{1}{N^3}$), cannot change the solution of equations
({\ref{Deq}) and ({\ref{Geq}) for $D > \frac{1}{N^{\frac{2}{3}}}$
and thus cannot lead to $q^* > \frac {1}{\sqrt{N}}$.  Therefore in
the mixed $p$-spin models we expect that higher order terms in
$\frac{1}{N}$ can only lead to power-law corrections in ${\cal P}$,
and thus do not qualitatively affect our results for 
$\frac{1}{N}\ln \frac {1}{{\cal P}}$.

\section{Discussion}

We have studied the attraction volume of a typical state
of energy $E$ in a family of disordered spherical spin models
which interpolate between $p=3$ (extensive configurational
entropy) and $p=2$ (one stable solution)
as a function of a tuning parameter $x$.  For $0 \le x < 1$ the total
number of metastable states is exponential in $N$.  We find that
for small $x$ (i.e. close to the $p=3$ model) the largest
attraction basin is an exponentially small fraction of the full
phase volume; this is true despite the fact that it is a strongly
varying function of energy.  We also find that for $x > x_c \sim 0.5$
the largest attraction basin constitutes a significant part of the
full phase volume, although the total number of states remains
exponential.
We did not find any thermodynamic signatures at $x = x_c$,
and thus believe that
{\sl only} the dynamical behavior of these glasses
changes qualitatively at this critical point. We note that the singularity
is approached as a function of decreasing $x$ with
increasing randomness in the models. Furthermore the
critical point
described here separating polynomially and exponentially small 
reduced attraction volumes $({\cal P} = \frac{W_B}{W_{PH}})$
as a function of $x$
bears striking ressemblance to
that studied recently in K-satisfiability problems.\cite{Monasson99}

An important open question is the physical origin of the state
with large attraction volume that appears for $1 > x \ge x_c$.  At
$x=1$ its presence is not surprising because there
exists only one stable solution.
It seems plausible that this 
state evolves continuously with decreasing
$x$ and retains its large attraction volume until $x=x_c$;
this has been confirmed by complementary numerical studies.
We denote this state by ${\cal A}(x)$.
Metastable states appear at $x < 1$; at values of $x$ just slightly
below 1 they have energies in a small interval $(E^*,E_m)$ separated
from $E_{\cal A}$, the energy of the state ${\cal A}$, by a gap
(cf. Figure 6) and exponentially small attraction basins. Thus,
$\cal{A}$ is both the optimal state
and the state with the largest attraction basin for $x$ close to
1.  In the limit $x \rightarrow 0$, ${\cal A}$ loses both of these special
features, namely it is no longer the ground-state and also has
an exponential small attraction basin.  Generically there are 
three possible ways that this can happen, shown schematically
in Figure 6.  Here we sketch the reduced attraction volume,
${\cal P} = \frac{W_B}{W_{PH}}$, as a function of $x$; for
$x < x_c$, ${\cal P}$ becomes exponentially small.  We also
show schematically the relative energy, $E_{\cal A} - E^*$, between
that of state ${\cal A}(x)$ and the lower edge of the ``continuous'' spectrum.
In scenario 1 (see Figure 6), ${\cal A}$ retains its optimal status in the vicinity
of $x_c$ even though it loses its large attraction volume.  By
contrast in case 3 (cf. Fig. 6), ${\cal A}$ loses first its optimal
character and then its large basin.  Finally in scenario 2 (cf. Fig. 6)
both special features are lost simultaneously; our complementary
numerical studies of the mixed $p=2$ and $p=3$ disordered spherical models
suggest that they are in this class.  
Solutions of optimization
problems that fall into category 1 (and perhaps category 2) may be 
accelerated
by noting that the ground-state for $x < x_c$ is continuously
connected to ${\cal A}_{x > x_c}$ by tuning the parameter
$x$. In principle one would start by locating ${\cal A}(x)$ for
$x > x_c$,
a relatively easy problem due to its large attraction
volume, and then reduce $x$ continuously to its value of interest
($x < x_c$).
This procedure is reminiscent of simulated annealing where
temperature plays the analogous role of the tuning parameter $x$.
It may therefore provide an alternative
optimization algorithm for a certain class of $NP$ problems.

We thank A. Barrat, A. Cavagna, S. Franz, I. Giardina, M. Mezard,
R. Zecchina
and particularly
A. Lopatin for useful discussions.

\section{Appendix I}

Here we sketch the derivation of the thermodynamic properties of the mixed 
$p=2$ and $p=3$ spherical spin models in the replica approach. 
Our main goal is to show that
low-temperature state is described by the one-step replica symmetry breaking
solution at \emph{all }$x$ such that $0 \le x < 1$. 
We follow the standard replica approach
developed for $p$-spin spherical models\cite{Crisanti92} with slight
modifications implied by the mixed case that we consider here. We introduce
order parameter $Q_{\alpha \beta }=\frac{1}{N}\sum_{i}\left\langle
S_{i,\alpha }S_{i,\beta }\right\rangle $ and integrate out the spin degrees
of freedom. We get the free energy as a function of $Q_{\alpha \beta }$
\begin{equation}
F(Q_{\alpha \beta })=-\frac{1}{4T}\left\{ \frac{1-x}{3}\sum Q_{\alpha \beta
}^{3}+\frac{x}{2}\sum Q_{\alpha \beta }^{2}\right\} -\frac{T}{2}\mathbf{Tr}%
\ln Q  \label{F(Q_ab)}
\end{equation}
which should be minimized over all $Q_{\alpha \beta }$ that satisfy the
constraint $Q_{\alpha \alpha }=1$. Varying this free energy with respect to $%
Q_{\alpha \beta }$ we get an expression for the order parameter
\begin{equation}
\frac{1}{4T}\left\{ \left( 1-x\right) Q_{\alpha \beta }^{2}+xQ_{\alpha \beta
}\right\} +\frac{T}{2}\widehat{Q}^{-1}{}_{\alpha \beta }=0  \label{Eq_Q_ab}
\end{equation}
where $\widehat{Q}^{-1}$ denotes matrix inversion. 

In order to
solve equation (\ref{Eq_Q_ab},
we multiply it by the $Q$ matrix, look for the solution in the form $\widehat{Q}=%
\widehat{1}+\widehat{q}$ and use the Parisi ansatz for 
the matrix $\widehat{q}$.
Next we exploit the structure of the matrices involved 
in order to solve the resulting equations in the limit 
$n\rightarrow 0$,
More specifically we note 
that two matrices, $A$ and $B$, that have the block structure of the Parisi
ansatz and are described by the functions $A(z)$ and $B(z)$ in the limit 
$n\rightarrow 0$ obey  the ''multiplication rule''
\begin{equation}
(\widehat{A}\widehat{B})_{z}=-\left[
\int_{0}^{z}A_{y}B_{y}dy+A_{z}\int_{z}^{1}B_{y}dy+B_{z}%
\int_{z}^{1}A_{y}dy+xA_{z}B_{z}\right] \label{Multiplication}
\end{equation}
Using this rule for matrices $A_{\alpha \beta }=\left( 1-x\right) q_{\alpha
\beta }^{2}+xq_{\alpha \beta }$ (i.e. $A_{z}=\left( 1-x\right)
q_{z}^{2}+xq_{z}$) and $B_{\alpha \beta }=q_{\alpha \beta }$, we see that for
all $z\neq 1$, i.e. for all non-diagonal elements of the corresponding
matrix, the (\ref{Multiplication}) becomes $-(\widehat{A}\widehat{B}%
)_{z}=(1-x)q_{z}^{2}+(1+x)q_{z}$. Differentiating this equation once with
respect to $z$, we obtain
\begin{equation}
A_{z}^{\prime }\int_{z}^{1}B_{y}dy+B_{z}^{\prime
}\int_{z}^{1}A_{y}dy+z(A_{z}^{\prime }B_{z}+A_{z}B_{z}^{\prime })=\left[
2(1-x)q_{z}+(1+x)\right] q_{z}^{\prime }
\end{equation}

All terms in the preceeding equation are proportional to $q^{\prime }$. 
Assuming that 
$q^{\prime }\neq 0$ (i.e. that solution is smooth), 
we divide it by $q^{\prime
}$ and differentiate it twice with respect to $z$.  We obtain 
the 
equation 
\begin{equation}
\frac{q^{\prime }}{4(1-x)q+2x}=-\frac{1}{6z(1-x)}  \label{Eq_q}
\end{equation}
that clearly does not allow a solution with positive $q$ and $q^{\prime }$.
Thus, we have proved that smooth solutions with $q^{\prime }\neq 0$
corresponding to continuous replica-symmetry breaking are {\sl impossible}.
Assuming now a one-step replica breaking corresponding to a step- like
function $q_{z}$ at $z=z_{0}$, we get the free energy
\begin{equation}
F(q)=+\frac{1}{4T}\left\{ \frac{1-x}{3}q^{3}+\frac{x}{2}q^{2}\right\} +\frac{%
T}{2z_{0}}\ln \frac{1-q}{1-(1-z_{0})q}-\frac{T}{2}\ln (1-q)  \label{F(q)}
\end{equation}
Numerical inspection of this function indicates that at low temperatures it
always has a maximum for some $0<z_{0}<1$ and $0<q<1$ corresponding to a
non-trivial one-step replica breaking solution.

\section{Appendix II}

We now present a skeletal derivation the number
of stable solutions, ${\cal N}(x,e)$,  
associated with the system of equations
\begin{eqnarray}
\lambda s_i &=& 2 \sum_j J_{ij} s_j + 3 \sum_{jk} J_{ijk} s_j
s_k\label{a1}\\
Ne &=&  \sum_{ij} J_{ij} s_i s_j + \sum_{ijk} J_{ijk}s_i s_j s_k.
\label{a2}
\end{eqnarray}
and the sum on the spin variables
$\sum_{i=1}^N s_i^2 = N$.
Here $e$ is the physical energy per spin of the metastable states, and
can be conveniently represented as a sum
\begin{equation}
e = \frac{1}{3} (\lambda + \epsilon)
\label{e}
\end{equation}
where $N\epsilon \equiv \sum_{ij} J_{ij} s_i s_j$ is the
energy contribution from the two-spin model.

In order to compute the number of metastable solutions,\cite{Tanaka80}
we use the expression
\begin{equation}
{\cal N}(\lambda,\epsilon,x) 
= \int \Pi_i \ ds_i \ \delta(\sum_i s_i^2 - N)
\ \delta \left( \frac{\partial H(\lambda)}{\partial s_i}\right )
\ \det \left( \frac{\partial H(\lambda)}{\partial s_i \partial
s_j}\right ) \ \delta(N\epsilon - \sum_{ij} J_{ij}s_i s_j)
\label{N(x,e)}
\end{equation}
where we perform the calculation at $T=0$, exploiting the absence
of subdivision of states for $T < T_c$.  
The determinant in (\ref{N(x,e)}) can be calculated by noting that
$A_{ij} \equiv \frac{\partial H(\lambda)}{\partial s_i \partial s_j}$
is a random symmetric matrix with a semicircular density of
eigenvalues distributed in the interval between $\lambda - \mu_0$
and $\lambda + \mu_0$.  We will see later that ${\cal N}(\lambda,
\epsilon)$
is dominated by $\lambda = \mu_0$; in this case 
\begin{equation}
{\cal D} = \ln (\det A_{ij}) = \frac{N}{2} \left( 1 + \ln
\frac{(2-x)}{2}\right)\label{det}
\end{equation}
where the $x$-dependence of the preceeding expression arises from 
disorder-averages over the couplings, $\langle J_{ij}^2\rangle$
and $\langle J_{ijk}^2 \rangle$.  Implementing an integral
representation
of the $\delta$-function, we write
\begin{equation}
{\cal N}(\lambda,\epsilon,x) = \int
\Pi_i \ ds_i \ \Pi_i \ \frac{d\phi_i}{2\pi} \  \frac{d \mu_i}{2\pi}
\ e^{i{\cal L} + {\cal D}} \ \delta(\sum_i s_i^2 - N)
\label{N2}
\end{equation}
where the effective Lagrangian is
\begin{equation}
{\cal L} = N\mu \epsilon + \lambda \sum_i \phi_i s_i - 
\mu \sum_{ij} s_i J_{ij} s_j - 2\sum_{ij} J_{ij}s_j\phi_i
- 3\sum_{ijk} J_{ijk}s_js_k\phi_i.
\label{l}
\end{equation}
We average the couplings over disorder to obtain
\begin{equation}
{\cal L} =  N\mu \epsilon + \lambda \sum_i \phi_i s_i
            + \frac{Ni}{4}\left\{ \frac{\mu^2 x}{2} + \frac{1}{N}
            \sum_i \phi_i^2 + \frac{(2-x)}{N^2} (\sum_i \phi_i s_i)^2
            + \frac{2\mu x}{N} \sum_i \phi_i s_i
            \right\}.
\label{l2}
\end{equation}
We note that the change of variables $\vec{\phi} \rightarrow
(\phi_{||},\phi_{\perp})$ where $\phi_{||} =
\frac{\phi_is_i}{\sqrt{N}}$ 
so that the effective Lagrangian is no longer a function of $s_i$.  We
can perform the integral over the spin variables in (\ref{N2})
with the result
\begin{equation}
{\cal N}(\lambda,\epsilon,x) = \int \frac{d\mu}{2\pi}
\ \frac{d\phi_{||}}{2\pi} \ \Pi_{\nu=1}^{N-1}
\ \frac{d\phi_{\perp}^\nu}{2\pi}
\ e^{i\tilde{\cal L} + \tilde{\cal D}}
\label{N3}
\end{equation}
where 
\begin{equation}
\tilde{\cal L} = N\mu\epsilon + \frac{iNx\mu^2}{8} +
\lambda\sqrt{N}\phi_{||}
+ \frac{1}{4} \left\{ 
(\phi^2_{||} + \phi^2_{\perp}) +
(2-x)\phi_{||}^2
+ 2 \mu x \sqrt{N}\phi_{||}\right\}
\label{Leff}
\end{equation}
and
\begin{equation}
\tilde{\cal D} = \frac{N}{2} \left\{ 2 + \ln \pi(2-x)\right\}.
\label{Deff}
\end{equation}      
Integrating over the remaining variables $\mu$, $\phi_{||}$ and
$\phi_{\perp}$ in (\ref{N3}) and using the relation
$\epsilon = 3e - \lambda$, we obtain
\begin{equation}
\ln {\cal N}(\lambda,e,x) = \frac{N}{2} \left \{ 2 + \ln (2-x) - 
\frac{2\lambda^2}{3-x} - \frac{12(e(x-3) +
\lambda)^2}{(3-x)x(1-x)}\right\}.
\label{N4}
\end{equation} 
In order to determine the total number of metastable states, we
maximize ${\cal N}(\lambda,x,e)$ with respect to $\lambda$ subject to
the constraint that $\lambda > \mu_0$ where
\begin{equation}
\mu_0 = 2\sqrt{N\langle A_{ij}^2\rangle} = \sqrt{2(2-x)}
\label{mu_0}
\end{equation}
to ensure that all eigenvalues of $A_{ij}$ are positive so that
we are only counting stable states.  We have checked that for 
$0 < x < 1$ the main contribution to ${\cal N}(\lambda,x,e)$ comes
form $\lambda = \mu_0$.  This implies that at any energy for $0 < x <
1$
the majority of the states are marginally stable.  In order
to obtain the total number of states, we maximize 
${\cal N}(x,e,\lambda=\mu_0)$
with respect to energy which yields the result (\ref{S(x)}).


\pagebreak

\begin{figure}
\centerline{\epsfxsize=10cm \epsfbox{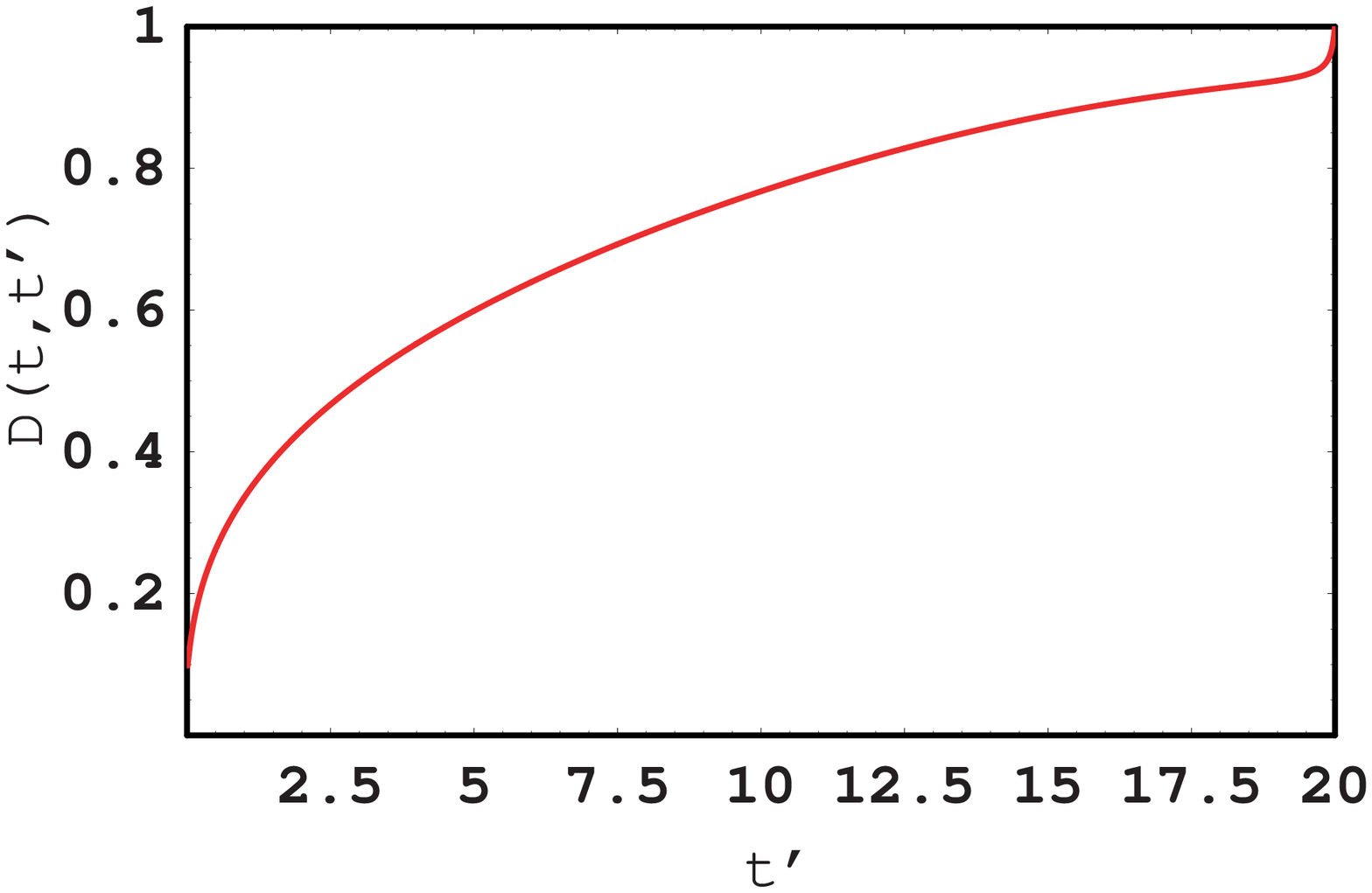}} {\footnotesize 
\textbf{Figure 1.}
Spin-spin correlation function after a fast quench at $T=\frac{1}{4}T_c$ for the $p=3$
spherical model.  Note the fast relaxation at very short $t - t' \ll 1$.}
\end{figure}

\begin{figure}
\centerline{\epsfxsize=10cm \epsfbox{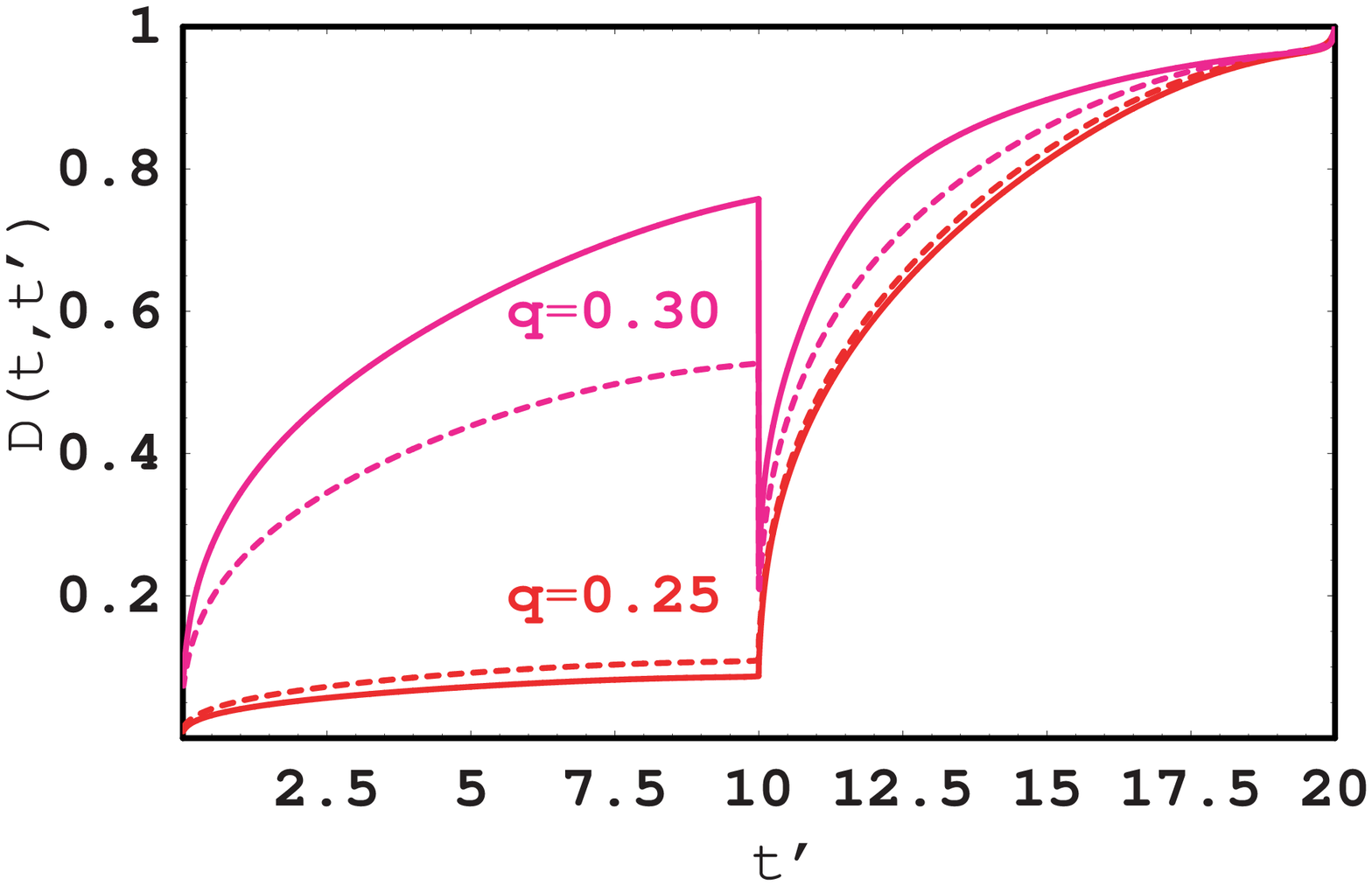}} {\footnotesize 
\textbf{Figure 2.}
The spin-spin   orrelation function for the $p=3$ spherical model after a
fast quench to $T = \frac{1}{8} T_c$ followed by a randomization
to $t_0 = \frac{1}{2} t$ with different fractions, $(1-q)$, of the
total number of spin affected.  Dashed lines refer to the results
for the total time $t=10$ rescaled to those for $t=20$.  This
data indicates that $0.25 < q^* < 0.30$ since an increase in the
total time $t$ leads to evolution in different directions
of $Q(q) \equiv \lim_{t\rightarrow\infty} D_{tt_0-\epsilon} (q)$.}
\end{figure}


\begin{figure}
\centerline{\epsfxsize=10cm \epsfbox{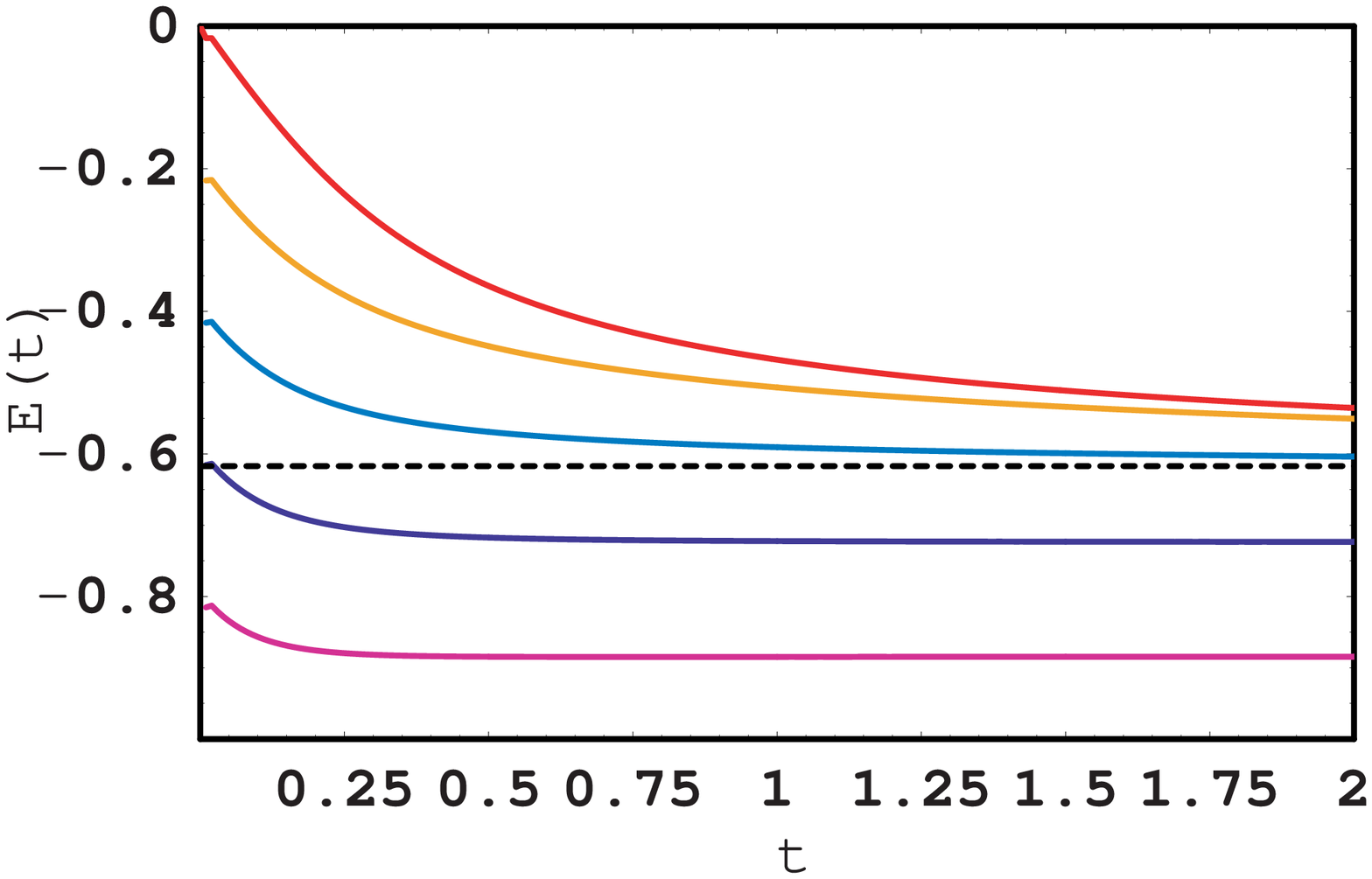}} {\footnotesize 
\textbf{Figure 3.}
The energy-dependence of the dynamical solutions for the $p=3$ spherical
model with different initial energies.  Note that for all initial
energies $E(0) > -0.4$ the energy approaches its asymptotic value
($E(\infty) = -0.61$, indicated by the dashed line) with power-law
decay; by contrast for $E_0 < -0.4$
the energy behavior is exponential and $E(\infty)$ depends on $E(0)$.}
\end{figure}


\begin{figure}
\centerline{\epsfxsize=10cm \epsfbox{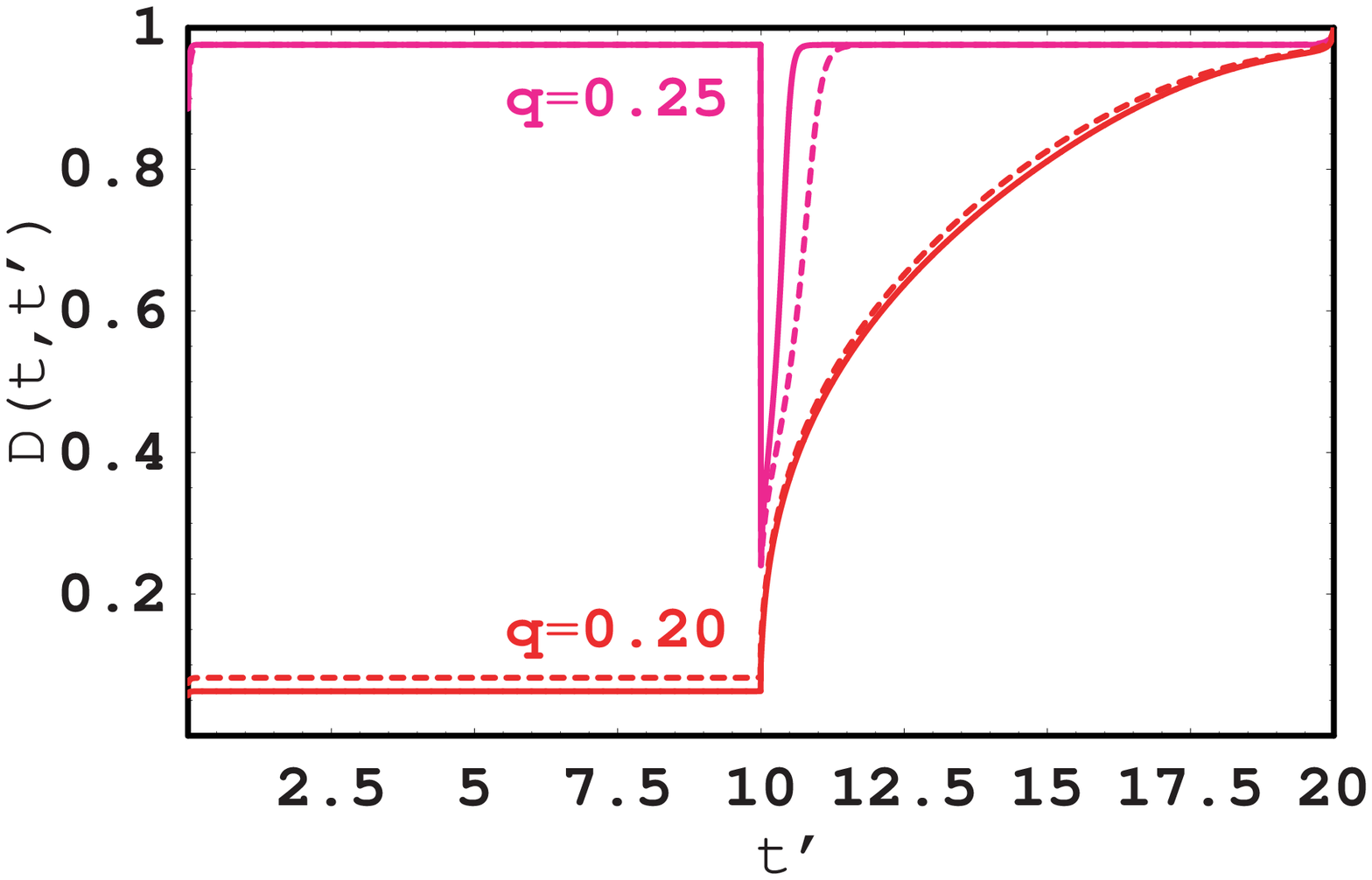}} {\footnotesize 
\textbf{Figure 4.}
The spin-spin correlation function for the $p=3$ model with 
$E(0) = -0.6$ (corresponding to exponential relaxation) with 
$T = \frac{1}{8} T_c$ with randomization at $t_0 = \frac{1}{2} t$.
Note that for $q = 0.25$ the randomization is followed
by very fast relaxation back to the initial state.  By contrast, a
slightly larger randomization $q= 0.20$ (corresponding to the 
randomization of $(1-q) = 0.80$  of the total spins) leads to
a completely different states similar to that found after
a fast quench.}
\end{figure}

\pagebreak

\begin{figure}
\centerline{\epsfxsize=10cm \epsfbox{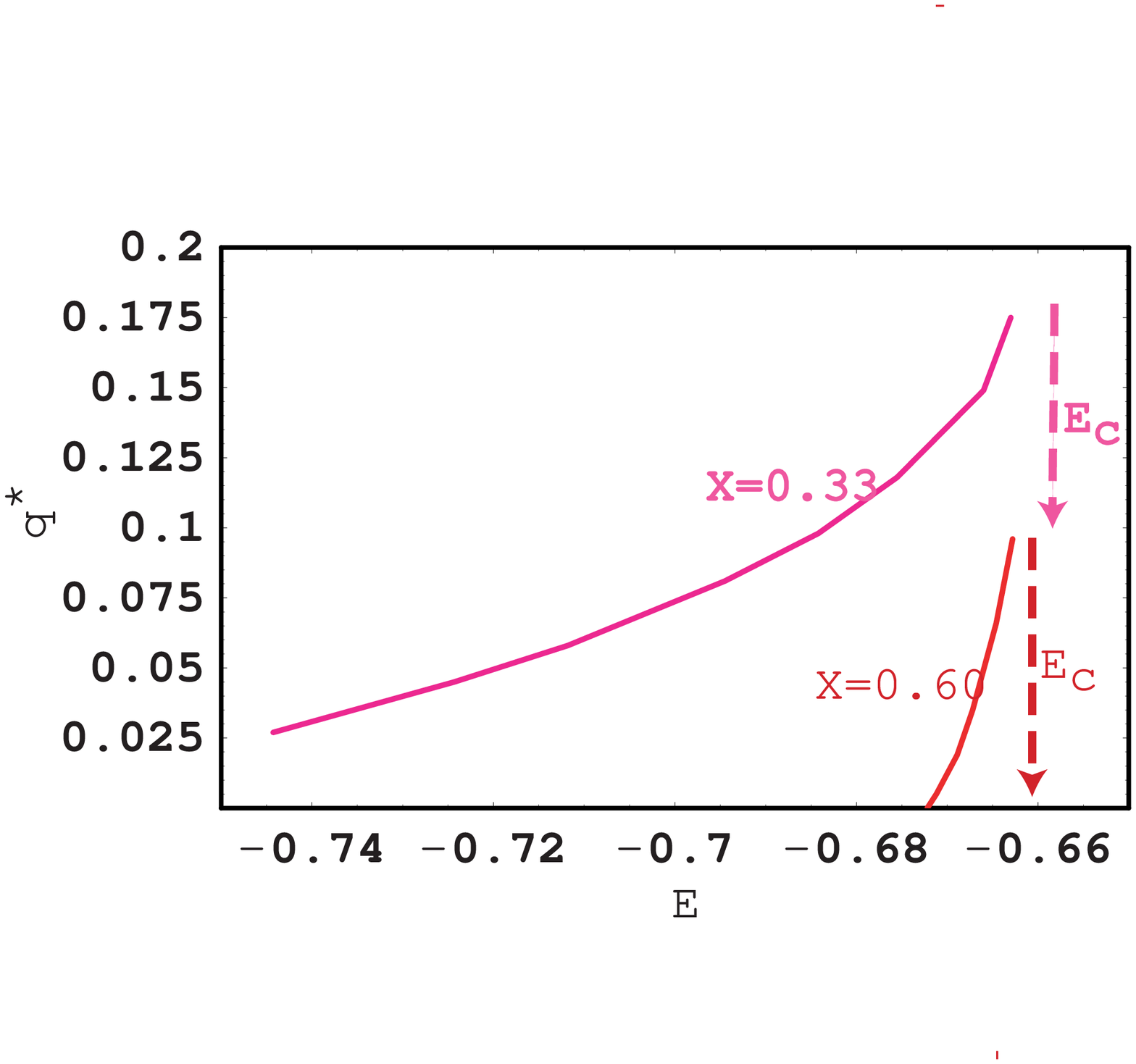}} {\footnotesize 
\textbf{Figure 5.}
The critical overlap, $q^*$, as a function of initial energy,
$E=E(0)$,
for $x=0.33$ ($x < x_c$) and $x=0.60$ ($x>x_c$).  We note that
in the latter case $q^*(E)$ crosses the x-axis, indicating the
appearance of a state with a large attraction volume.}
\end{figure}

\begin{figure}
\centerline{\epsfxsize=10cm \epsfbox{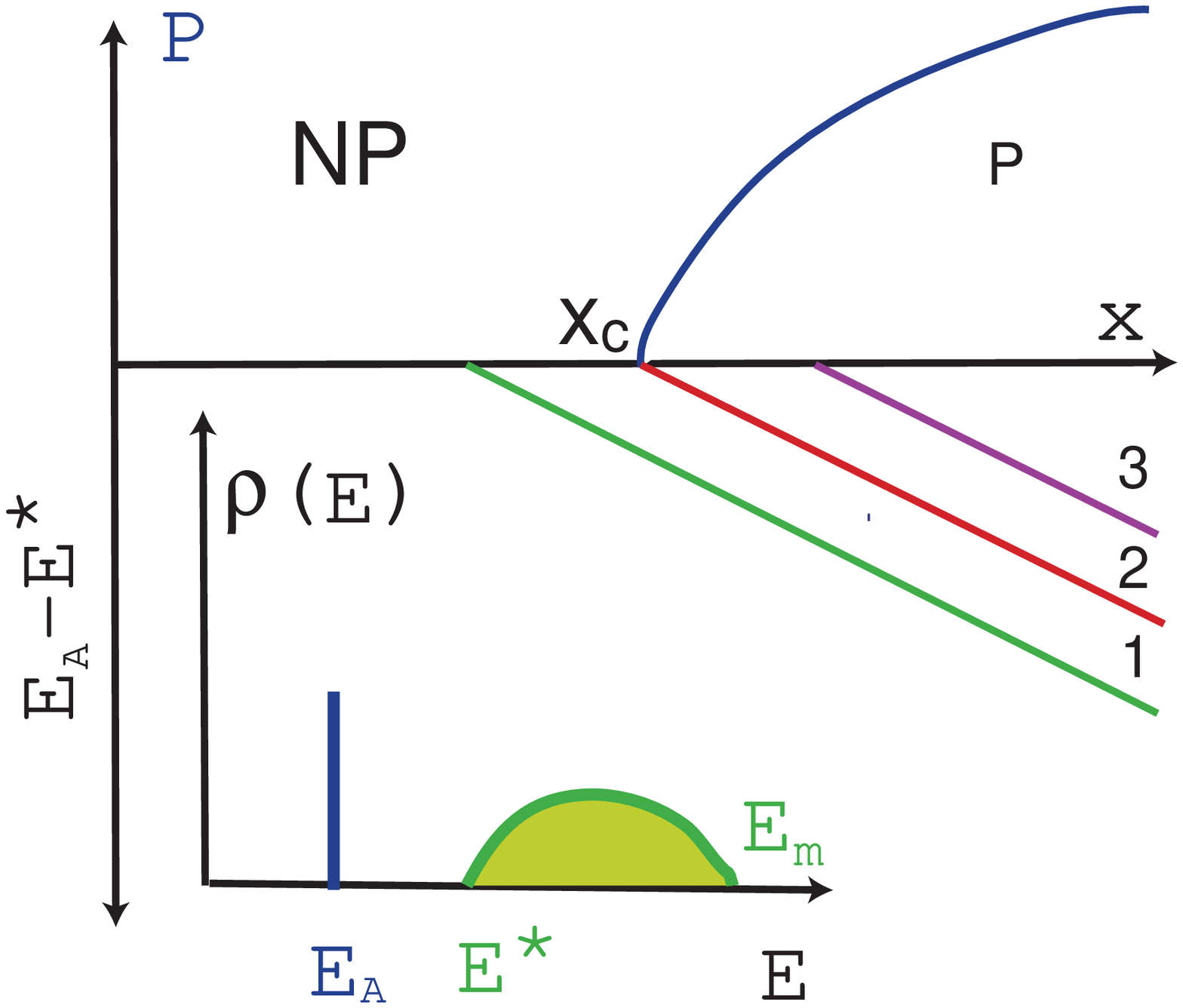}} {\footnotesize 
\textbf{Figure 6.}
A schematic of the reduced typical attraction volume, 
${\cal P} = \frac {W_B}{W_{PH}}$, 
and the relative energy, $E_{\cal A} - E^*$,
as a function of $x$; the three scenarios for $E_{\cal A} - E^*$
in the approach to $x_c$ are described in the text.}

\end{figure}

\end{document}